\documentclass[notitlepage, twocolumn, 9pt]{article}
\usepackage[a4paper]{geometry}
\usepackage[parfill]{parskip}
\usepackage{enumitem}
\usepackage[T1]{fontenc}
\usepackage[utf8]{inputenc}
\usepackage{graphicx}
\usepackage{amsmath}
\usepackage{pgfplots}
\usepackage{textcomp}
\usepackage{fancyhdr}
\usepackage{setspace}
\usepackage[justification=centering]{caption}
\usepackage{xcolor}
\usepackage{framed}
\usepackage{sectsty}
\usepackage[colorlinks, plainpages=false, pdfpagelabels]{hyperref}
\usepackage{tikz}
\usepackage{adjustbox}
\usepackage{caption}
\usepackage{multirow}
\usepackage{array}
\DeclareGraphicsExtensions{ .jpg, .png, .pdf}

\parskip = \baselineskip

\pgfplotsset{
  width=10cm,
  compat=1.9,
}

\usetikzlibrary{positioning, arrows}
\usetikzlibrary{decorations.pathreplacing}

\hypersetup{
  colorlinks,
  linktoc=all,
  urlcolor=blue,
  linkcolor=black,
  citecolor=blue
}

\setlist{parsep=0.5em}

\let\OLDthebibliography\thebibliography
\renewcommand\thebibliography[1]{
    \OLDthebibliography{#1}
    \setlength{\parskip}{0pt}
    \setlength{\itemsep}{8pt plus 1ex}
}

\begin{document}
\twocolumn[{
\begin{centering}
  \begin{spacing}{2}
    {\huge \textsc{CHRONOMID}} \\
    \vskip 1.2mm
    {\LARGE \textsc{cross-modal neural networks for }}
    {\Large \textsc{3}}%
    {\LARGE \textsc{-d temporal medical imaging data}
    }
  \end{spacing}
  \begin{spacing}{1.3}
    {\textit{Alexander G. Rakowski, Petar Veli\v{c}kovi\'c, Enrico Dall'Ara, Pietro Li\`o}} \\
    \vspace{10mm}
  \end{spacing}
\end{centering}
}]
%
\begin{abstract}
ChronoMID builds on the success of cross-modal convolutional neural networks
(X-CNNs), making the novel application of the technique to medical imaging data.
Specifically, this paper presents and compares alternative approaches ---
timestamps and difference images ---
to incorporate temporal information for the classification of bone disease in mice,
applied to \textmu CT scans of mouse tibiae.
Whilst much previous work on diseases and disease classification has been based
on mathematical models incorporating domain expertise and the explicit encoding
of assumptions, the approaches given here utilise the growing availability of
computing resources to analyse large datasets and uncover subtle patterns
in both space and time.
After training on a balanced set of over 75000 images,
all models incorporating temporal features outperformed a state-of-the-art
CNN baseline on an unseen, balanced validation set comprising over 20000 images.
The top-performing model achieved 99.54\% accuracy, compared to 73.02\% for
the CNN baseline.
\end{abstract}
%
%
\section{Introduction}
Temporal information is especially important in the medical domain,
wherein diseases develop over extended periods,
treatment efficacy cannot be evaluated at the moment of administration,
and temporal proximity between diseases suffered by a patient affects
treatment suitability and patient recovery.
In particular, studying bone remodelling
(i.e. the changes over time of bone properties regulated by the activity of the bone cells)
is fundamental for better understanding the effect of musculoskeletal diseases,
such as osteoporosis and osteoarthritis, and to optimise the related treatments.
Animal models, in particular mice, are used for studying the effect of new treatments
on bone remodelling using high-resolution medical images and
micro computed tomography (\textmu CT).
From \textmu CT images, proper assessment of the morphometric parameters
of the bone microstructure can be used to measure the effect of
diseases or interventions \cite{bouxsein10}.
Furthermore, 3D \textmu CT images collected in vivo can be converted into
biomechanical computational models
(i.e. finite element models based on partial differential equations)
for the non-invasive assessment of the bone's mechanical properties \cite{oliviero18}.
Medical data, medical imaging data especially, contains a wealth of information
which could, for example, inform diagnoses or detect early-stage diseases,
yet remains largely untapped by automated processes.
Due to the sheer quantities of data involved in many medical problems
and the potential subtlety of informative patterns within it,
the task of uncovering and extracting such patterns often falls outside the scope of
human capabilities.
Data-driven, automated approaches are essential to deriving usable insights from
large medical datasets, such as those generated by scanners or ECGs.
Prior work to incorporate temporal information for healthcare has focused on
mining electronic patient health records for elucidating textual content or
data in structured fields, thereby neglecting the temporal aspects of imaging
data \cite{choi16}.
In this paper, we build on the successes of cross-modal convolutional neural networks
\cite{velickovic16} to incorporate both spatial and temporal information explicitly into models.
We present five models, each using an alternate description of temporal information, and compare
their performance against a state-of-the-art conventional convolutional neural network (CNN)
for the task of bone disease classification in mice.
Our approaches involved relative timestamps, corresponding to the week the scan was taken since
the study began, absolute and relative difference images, and the combination of timestamps and
difference images.
Two groups of mice were studied: a control group of healthy mice and a group
that were treated with parathyroid hormone (PTH),
which can have anabolic effects leading to accelerated bone formation
when administered correctly.
The data for these mice --- treatment status and \textmu CT scans of
the right tibia of each mouse --- were provided by
Dr. Enrico Dall'Ara's team at the University of Sheffield's
Department of Oncology and Metabolism.
Whilst different variants of recurrent neural networks (RNNs) are increasingly common for tasks
involving sequences of data, they have limitations and drawbacks we seek to overcome.
Foremost, RNNs are designed to process sequential inputs, whereas medical data may have missing
data points, e.g. due to patients missing appointments, or data points provided out of order,
e.g. due to different administrative bodies holding patient records because people move locations
or switch healthcare providers.
It is preferable to make diagnoses as data becomes available rather than waiting
until it has been provided in its entirety.
Furthermore, RNNs are computationally expensive, partially due to the multiple gates and units
involved in memory cells, and partially due to intra-layer or backwards as well as forwards
connections, leading to potentially very many parameters,
in turn increasing the amount of data required to train, thus training times.
Moreover, by decoupling the spatial and temporal information of the image set,
we allow the networks to determine the relative importance of each component.

The techniques we developed are expected to perform well with sparse data sets
as a consequence of previous findings on X-CNNs and the representations of time used,
which permit potentially large and irregular gaps between data points.
Our results are encouraging from both a computational perspective and a
medical one: incorporating even simple descriptions of time into models can yield notable
improvements, with more suitable temporal descriptions producing significantly
better classification performance.

%
%
%
%
\section{Medical Background}

\newcommand{\mousebonefullheight}{15mm}
\newcommand{\mousebonevertsep}{3mm}
\newcommand{\mousebonehorizsep}{5mm}
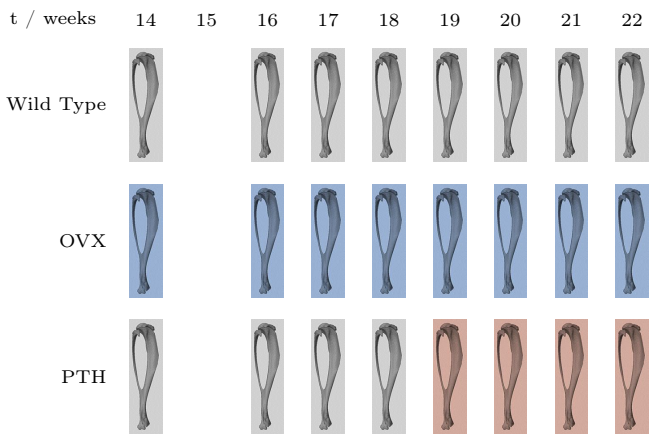
\begin{figure}[!b]
  \centering
  \begin{tikzpicture}[
      every node/.style={
        inner sep=0mm,
        outer sep=0mm,
        font=\scriptsize,
      },
      mousewild/.style={
        rectangle,
        minimum height=\mousebonefullheight,
        label={center:\includegraphics[height=\mousebonefullheight]{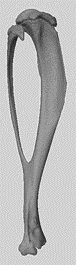}}
      },
      mouseovx/.style={
        rectangle,
        minimum height=\mousebonefullheight,
        label={center:\includegraphics[height=\mousebonefullheight]{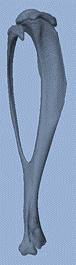}}
      },
      mousepth/.style={
        rectangle,
        minimum height=\mousebonefullheight,
        label={center:\includegraphics[height=\mousebonefullheight]{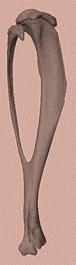}}
      },
      dummy/.style={
        fill=none,
        minimum width=3mm,
      }
    ]
    \node[dummy] (d1) {14};
    \node[dummy] (d2) [right=\mousebonehorizsep of d1] {15};
    \node[dummy] (d3) [right=\mousebonehorizsep of d2] {16};
    \node[dummy] (d4) [right=\mousebonehorizsep of d3] {17};
    \node[dummy] (d5) [right=\mousebonehorizsep of d4] {18};
    \node[dummy] (d6) [right=\mousebonehorizsep of d5] {19};
    \node[dummy] (d7) [right=\mousebonehorizsep of d6] {20};
    \node[dummy] (d8) [right=\mousebonehorizsep of d7] {21};
    \node[dummy] (d9) [right=\mousebonehorizsep of d8] {22};
    \node[mousewild] (w1) [below=\mousebonevertsep of d1] {};
    \node[mousewild] (w3) [below=\mousebonevertsep of d3] {};
    \node[mousewild] (w4) [below=\mousebonevertsep of d4] {};
    \node[mousewild] (w5) [below=\mousebonevertsep of d5] {};
    \node[mousewild] (w6) [below=\mousebonevertsep of d6] {};
    \node[mousewild] (w7) [below=\mousebonevertsep of d7] {};
    \node[mousewild] (w8) [below=\mousebonevertsep of d8] {};
    \node[mousewild] (w9) [below=\mousebonevertsep of d9] {};
    \node[mouseovx] (o1) [below=\mousebonevertsep of w1] {};
    \node[mouseovx] (o3) [below=\mousebonevertsep of w3] {};
    \node[mouseovx] (o4) [below=\mousebonevertsep of w4] {};
    \node[mouseovx] (o5) [below=\mousebonevertsep of w5] {};
    \node[mouseovx] (o6) [below=\mousebonevertsep of w6] {};
    \node[mouseovx] (o7) [below=\mousebonevertsep of w7] {};
    \node[mouseovx] (o8) [below=\mousebonevertsep of w8] {};
    \node[mouseovx] (o9) [below=\mousebonevertsep of w9] {};
    \node[mousewild] (p1) [below=\mousebonevertsep of o1] {};
    \node[mousewild] (p3) [below=\mousebonevertsep of o3] {};
    \node[mousewild] (p4) [below=\mousebonevertsep of o4] {};
    \node[mousewild] (p5) [below=\mousebonevertsep of o5] {};
    \node[mousepth] (p6) [below=\mousebonevertsep of o6] {};
    \node[mousepth] (p7) [below=\mousebonevertsep of o7] {};
    \node[mousepth] (p8) [below=\mousebonevertsep of o8] {};
    \node[mousepth] (p9) [below=\mousebonevertsep of o9] {};
    \node[dummy] (ltime) [left=\mousebonehorizsep of d1] {t / weeks};
    \node[dummy] (lwild) [left=\mousebonehorizsep of w1] {Wild Type};
    \node[dummy] (lovx) [left=\mousebonehorizsep of o1] {OVX};
    \node[dummy] (lpth) [left=\mousebonehorizsep of p1] {PTH};
  \end{tikzpicture}
  \caption{Treatment timeline of mouse groups, where $t$ is the age of the mouse in weeks.
    Grey indicates no treatment, blue indicates ovariectomy (OVX),
    and red indicates treatment with PTH.
  }
  \label{fig-mouse-treatments}
\end{figure}

\newcommand{\mousebonexsectionheight}{30mm}
\newcommand{\mousebonexsectionsep}{15mm}
\begin{figure}
  \centering
  \begin{tikzpicture}[
      every node/.style={
        font=\scriptsize,
        fill=none,
        inner sep=0mm,
        outer sep=0mm,
        minimum height=\mousebonexsectionheight,
      },
      dummy/.style={
        minimum height=0mm,
      },
      connector-between/.style={
        red,
        dashed,
        -,
        very thick,
      },
      connector-at/.style={
        red,
        ->,
        very thick
      },
    ]
    \node (bone-full)
      {\includegraphics[height=\mousebonexsectionheight]{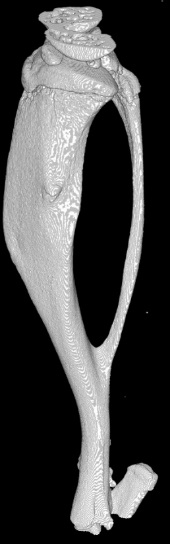}};
    \node (bone-plates) [right=\mousebonexsectionsep of bone-full]
      {\includegraphics[trim=0pt 30pt 17pt 36pt,
        clip=true,
        height=\mousebonexsectionheight]{mouse-bone-full.jpg}};
    \node (bone-slices) [right=\mousebonexsectionsep of bone-plates]
      {\includegraphics[height=\mousebonexsectionheight]{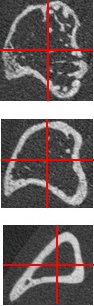}};
    \node[dummy] (bone-full-l1) [below=6mm of bone-full.north west] {};
    \node[dummy] (bone-full-r1) [below=6mm of bone-full.north east] {};
    \node[dummy] (bone-full-l2) [above=6mm of bone-full.south west] {};
    \node[dummy] (bone-full-r2) [above=6mm of bone-full.south east] {};
    \node[dummy] (bone-plates-l1) [below=4mm of bone-plates.north west] {};
    \node[dummy] (bone-plates-r1) [below=4mm of bone-plates.north east] {};
    \node[dummy] (bone-plates-l2) [below=12mm of bone-plates.north west] {};
    \node[dummy] (bone-plates-r2) [below=12mm of bone-plates.north east] {};
    \node[dummy] (bone-plates-l3) [below=18mm of bone-plates.north west] {};
    \node[dummy] (bone-plates-r3) [below=18mm of bone-plates.north east] {};
    \node[dummy] (bone-slices-top) [below=5mm of bone-slices.north west] {};
    \node[dummy] (bone-slices-mid) [below=16mm of bone-slices.north west] {};
    \node[dummy] (bone-slices-bot) [below=26mm of bone-slices.north west] {};
    \path[connector-between] (bone-full-l1) edge (bone-full-r1);
    \path[connector-between] (bone-full-r1) edge (bone-plates.north west);
    \path[connector-between] (bone-full-l2) edge (bone-full-r2);
    \path[connector-between] (bone-full-r2) edge (bone-plates.south west);
    \path[connector-between] (bone-plates-l1) edge (bone-plates-r1)
	  (bone-plates-l2) edge (bone-plates-r2)
  	  (bone-plates-l3) edge (bone-plates-r3);
    \path[connector-at] (bone-plates-r1) edge (bone-slices-top)
 	  (bone-plates-r2) edge (bone-slices-mid)
      (bone-plates-r3) edge (bone-slices-bot);
  \end{tikzpicture}
  \caption{Cross-sections of a mouse tibia between the proximal and distal growth plates.
    Left: full tibia.
    Middle: section of interest between the growth plates.
    Right: cross-sections at three locations along the tibia.
  }
  \label{fig-bone-cross-sections}
\end{figure}

There has been growing interest in automated disease diagnosis
and assessment based on machine learning (ML) models.
This interest has focused on extracting features from time-series,
such as using EEG data to predict epileptic attacks
up to an hour in advance \cite{guler05},
and on various types of medical imaging.
In \cite{ashinsky17}, Ashinsky et al seek to predict osteoarthritis using
images from MRI scans, whilst in \cite{cupek16}, Cupek et al worked on the
automated assessment of joint synovitis using ultrasound images,
applying a set of image processing and ML techniques to determine synovitis severity.
However, these two areas of interest have typically been studied separately,
thereby potentially excluding relevant data, features, and modelling techniques
which could improve model performance.

The case study explored in this paper concerns bone remodelling
and the classification of imbalanced remodelling.
Remodelling encompasses bone changes due to tissue maintenance
and is governed by two interacting mechanisms: formation and resorption,
in turn dependent on the RANK-RANKL-OPG signalling pathway \cite{pivonka10}.
Everyday actions, such as walking, cause micro-cracks in bone to develop.
If this micro-damage were simply filled in with mineralised extracellular matrix,
the old damaged tissue would continue to be weak, compromising the mechanical integrity
of the bone and risking further micro-damage accumulation.
Resorption is the process of removing the damaged tissue,
triggered by osteocytes signalling and performed by osteoclasts.
Osteoblasts prevent excessive bone excavation by producing a decoy receptor, OPG,
as they mature, inhibiting osteoclastic activity.
The osteoblasts then deposit collagen and induce mineralisation,
thereafter differentiating themselves as osteocytes embedded in the extracellular matrix.
In healthy bone tissue, these processes are coupled such that the site of the micro-crack
is repaired completely.
However, pathologies can cause imbalances that affect bone remodelling.
Osteoarthritis and osteopetrosis cause excessive mineralisation,
typically affecting joints due to these experiencing the greatest impacts
and general wear-and-tear.
This can ultimately lead to inflammation of synovial tissues,
restricting movement and paining the sufferer.
In contrast, diseases such as osteoporosis and osteomyelitis are caused by
excessive osteoclastic activity gradually degrading bone density and
reducing structural integrity,
in turn making the damaged bone more vulnerable to further micro-cracks,
causing additional deterioration in a vicious cycle.
The standard approach to modelling the coupled processes of formation and resorption
is to use stochastic simulations based on ordinary different equations or,
when extending temporal models to spatio-temporal ones,
partial differential equations.
These are important for understanding systems and
can use formal and probabilistic checking to verify simulation properties,
but are limited by the complexity of extending such models
and their inability to be used in automated data-processing tasks.


\section{Mouse Data}

The data were \textmu CT scans of the tibiae of
15 mice taken over an eight-week period, between the pages of 14 and 22 weeks,
which were used to study longitudinal bone changes caused by
the processes of bone resorption and formation
\cite{lu17}.
The 15 female mice were divided into three groups ($N=5$ per group),
as shown in Figure \ref{fig-mouse-treatments},
including a healthy control group and two groups with induced imbalanced
bone remodelling.
The control group were referred to by the researchers as being \emph{wild} type.
The bone formation group was treated with parathyroid hormone (PTH) ---
an anabolic peptide ---
four weeks into the experiment, at 18 weeks of age.
As such, from ages 14 to 18 weeks this group was treated identically to the control group.
The bone resorption group underwent ovariectomy at 14 weeks of age,
causing \oe strogen deficiency and accelerating bone resorption in the weeks 14 to 22,
particularly localised in the most proximal part of the bone, close to the knee.
Only female mice were considered for this reason,
as observed differences could be attributed to treatments rather than
potentially being the result of the sex of the mice in the different groups.
For our case study, we considered data from only the \textit{wild} type and \textit{PTH} groups.

\begin{figure}
  \begin{minipage}{0.5\linewidth}
    \includegraphics[width=0.95\textwidth]{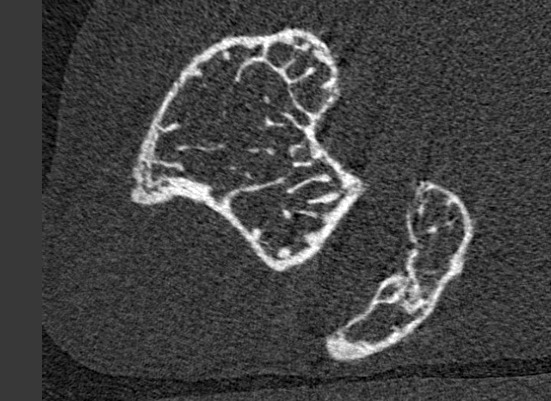}
    \hfill
  \end{minipage}%
  \begin{minipage}{0.5\linewidth}
    \hfill
    \includegraphics[width=0.95\textwidth]{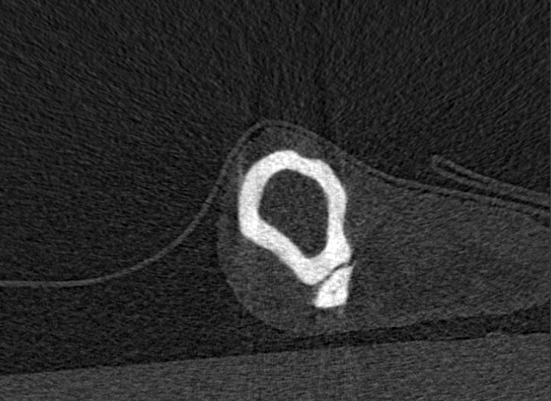}
  \end{minipage}
  \caption{\textmu CT scan images of the right tibia of a female mouse,
    taken at different longitudinal positions along the bone.
    Left: closer to the proximal epiphysis, near the knee.
    Right: closer to the distal epiphysis, near the foot.
    Brighter areas indicate denser tissue.
  }
  \label{fig-mouse-dicom}
\end{figure}

In order to monitor bone changes over space and time,
weekly \textmu CT scans of the right tibia of each mouse were taken.
These high-resolution scans were reconstructed in a series of slices:
cross-sectional images perpendicular to the axis of the tibia,
as depicted in Figure \ref{fig-bone-cross-sections},
with a voxel size of 10\textmu m in between the growth plates of the tibiae.
This resulted in around 1200-1400 slices per mouse per week,
giving a total of around 52 000 images with a combined size of
around 20GiB per test group.
Two images, both from the same mouse at the same time,
but at different longitudinal positions,
are given in Figure \ref{fig-mouse-dicom}.
Dr Dall'Ara's research team performed image registration to place the images
from different mice and different time points in the same reference system
\cite{oliviero17}.
Due to the use of an operator crop to remove regions which did not include the
tibia from the images generated by the \textmu CT scanner,
and in part due also to the mice still growing slightly,
despite being considered skeletally mature,
the image dimensions differ slightly from week to week and from mouse to mouse.
This variation in dimensions is characterised by image heights between
400-500 pixels and widths between 500-700 pixels.
All images were provided and stored in DICOM
(Digital Imaging and Communications in Medicine) format.
Many image processing libraries cannot accommodate variable image size,
and the data were incomplete:
no data were recorded for any of the mice for week two of the experiment,
and one of the mice in the induced bone formation (anabolic) group was missing
data for the final week.
Furthermore, the data exhibited a issue common in medicine and bioinformatics:
being deep rather than wide.

%
%
\section{Preprocessing}
\label{sec-preprocessing}

In order to both accommodate the differing image sizes and week-to-week alignment
variations, and to produce the different temporal descriptions, a number of
pre-processing steps were used.
The pipeline of these steps is shown in Figure \ref{fig-pipeline}.

As most image processing libraries require images with consistent dimensions,
it was necessary to convert all images to have the same width and height.
As the full dataset was known in advance, it was possible to standardise the
dimensions by expanding each image to the maximum height and width values
observed in the dataset, which were 501 and 763 pixels respectively.
This expansion process placed the image centrally and set any surrounding pixels
to the minimal intensity value in that image.
When the maximal image dimensions are not known in advance, images could be
up- or down-sampled to fit within pre-set dimensions,
either preserving relative dimensions and filling empty pixels as described,
or by fitting the image to the pre-set dimensions without preserving relative sizes.
This resizing process was performed once and resulting images saved to disk
to avoid the cost of resizing during each training phase.
In order to perform differencing between any two expanded images,
it was first necessary to align them, as the mouse from which the images were taken
would not be in exactly the same position from week to week.
The alignment was performed using \texttt{scikit-image}'s registration method,
which for a 2D image outputs a pair of numbers indicating the $x$ and $y$ axis
translations to perform.
The first image underwent the specified translation for optimal alignment with the
second image, followed by pixel-wise subtraction to create the difference map.

When each image was processed, an integer timestamp was recorded in the image meta-data,
representing the week the image was taken in since the experiment start.
Integer week numbers were recorded rather than normalised values for generalisability
beyond the dataset of the case study.
Timestamps normalised to the range [0,1], for example, this would require
experiment durations to be known in advance, thereby excluding later data.
Such timestamps would furthermore be incompatible between datasets which,
whilst otherwise comparable, had been collected over different durations.

The final stage of preprocessing was to split the dataset into three partitions
for training, validation, and testing.
The data for one mouse of the five in each treatment group was selected at random
and held out solely for testing.
The remaining data was randomly permuted and partitioned as $90\%$ for training and
$10\%$ for validation.
Random permutation helps to avoid biases from repeated exposure of a network
to similar training sequences and their associated gradient updates,
thereby mitigating the risk of overfitting.

Due to the nature of the images, some common preprocessing steps were either
undesirable or unnecessary.
As the \textmu CT scans were monochromatic, there was no opportunity to split the
images into multiple colour channels or use intensity-chromacity encodings.
Separating the differing information contained in these image channels can
boost model performance \cite{velickovic16},
thus, where applicable, chromacity and/or intensity channels
should be incorporated into medical X-CNNs,
applied to both the original images and the difference images.
Doing so would require only minor modifications to the X-CNN approaches presented herein.
In order to prevent compression artifacts, no compression was applied to the images.
Instead, the neural networks effectively perform sub-sampling via pooling,
in the process learning weights;
sub-sampling external to the network removes learning opportunities for the network itself.
Finally, data augmentation - the process of synthesising novel datasets via
some perturbation or transformation of an existing dataset - was not applicable.
Augmentation is fundamentally incompatible with differencing approaches
as the synthesised images are generated independently from one another.
This means there is no guaranteed correlation in the differences
between a reference-comparison image pair,
so performing differencing on augmented images
would amount to differencing randomly-selected image pairs,
thereby invalidating the intent of differencing.
Furthermore, augmentation is frequently used to expand small datasets
and ``fill in'' sparse ones.
For the dataset in this case study, there were large quantities of data for each mouse -
around $10 000$ images - and as the \textmu CT process produced image slices with high spatial proximity,
adjacent images should be highly correlated and with minimal variation.
Effectively, adjacent image slices appear to be perturbed, noisy versions of one another
and week-to-week alignment variations effectively introduce small translations and rotations,
both whilst retaining the temporal relationships between image slices of the
same individuals from different timestamps.
Thus, augmentation would be unlikely to benefit this and similar datasets.

\newcommand{\pipelinehorizsep}{3mm}
\newcommand{\pipelinevertseplarge}{20mm}
\newcommand{\pipelinevertsepsmall}{3mm}
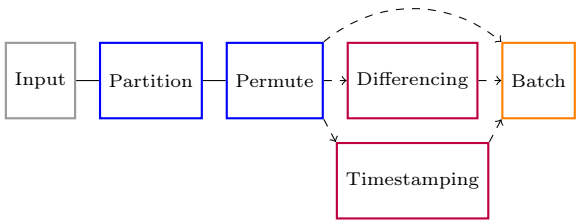
\begin{figure}[!ht]
  \begin{center}
    \begin{tikzpicture}[
      every node/.style={font=\scriptsize},
      inout/.style={rectangle, draw=black!40, fill=none, thick,
        minimum height=10mm, node distance=2.0cm,
      },
      proc/.style={rectangle, draw=blue, fill=none, thick,
        minimum height=10mm, node distance=2.0cm,
      },
      option/.style={rectangle, draw=purple, fill=none, thick,
        minimum height=10mm, node distance=2.0cm,
      },
      gen/.style={rectangle, draw=orange, fill=none, thick,
        minimum height=10mm, node distance=2.0cm,
      },
      ml/.style={rectangle, draw=red, fill=none, thick,
        minimum height=10mm, node distance=2.0cm,
      },
    ]
      \node[inout]    (in) {Input};
      \node[proc]     (part)    [right=\pipelinehorizsep of in]    {Partition};
      \node[proc]     (perm)    [right=\pipelinehorizsep of part]  {Permute};
      \node[option, align=center]
                      (diff)    [right=\pipelinehorizsep of perm]  {Differencing};
      \node[option]   (tstamp)  [below=\pipelinevertsepsmall of diff]  {Timestamping};
      \node[gen, align=center]
                      (gen)     [right=\pipelinehorizsep of diff]  {Batch};
      \path[->, dashed]
                      (perm.east) edge (diff.west)
                      (perm.south east) edge (tstamp.north west)
                      (diff.east) edge (gen.west)
                      (tstamp.north east) edge (gen.south west);
      \path[->, dashed, bend left=40]
                      (perm.north east) edge (gen.north west);
      \path[-]        (in.east) edge (part.west)
      				  (part.east) edge (perm.west);
    \end{tikzpicture}
  \end{center}
  \caption{Labelled images are partitioned into training, validation,
    and test datasets, then permuted to inject randomness.
    Temporal information was optionally incorporated via timestamps,
    difference images, or both;
    other descriptions of time may be substituted.
  }
  \label{fig-pipeline}
\end{figure}
\vspace{-1em}
%
%
\section{Models}

We present three different temporal descriptions and two combinations thereof,
compared against a state-of-the-art CNN baseline.
Whilst the architectures for each approach differ, they share common elements.
All models used a batch size of 32, employed max-pooling, batch normalisation,
and dropout, and utilised rectified linear unit (ReLU) activation functions
for their strong performance with convolutional networks \cite{nair10}.
We chose He initialisations for suitability with ReLU activations \cite{he15},
as the Glorot initialisation scheme on which they are based, whilst generally applicable,
was designed for sigmoidal activations {\cite{glorot10}}.
The output layer of each network was identical: one neuron per treatment class,
values normalised by softmax, with categorical cross-entropy as the loss function.

\begin{figure}[!ht]
  \begin{center}
  	\begin{tikzpicture}[
        every node/.style={font=\scriptsize},
        chain/.style={rectangle, draw=blue!100, fill=none, thick, minimum size=5mm,},
        conv/.style={rectangle, draw=orange, fill=none, thick, minimum size=5mm,},
        merge/.style={rectangle, draw=purple, fill=none, thick, minimum size=5mm,},
      ]
      \node[chain] (chain1) {Chain};
      \node[chain] (chain2) [below=5 mm of conv1] {Chain};
      \node[conv] (x-conv1) [right=5 mm of chain1] {Conv2D};
      \node[conv] (x-conv2) [right=5 mm of chain2] {Conv2D};
      \node[merge] (merge1) [right=5 mm of x-conv1] {Merge};
      \node[merge] (merge2) [right=5 mm of x-conv2] {Merge};
      \node[align=right](ti) [left=7mm of chain1] {Image,\\ time $t$};
      \node[align=right](td) [left=7mm of chain2] {Difference Image,\\ time $t_0$};
      \path[->] (chain1) edge (x-conv1)
      			(chain2) edge (x-conv2)
                (x-conv1) edge (merge1)
                (x-conv2) edge (merge2)
                (x-conv1.east) edge (merge2.west)
                (x-conv2.east) edge (merge1.west);
      \path[->, bend left=30] (chain1) edge (merge1);
      \path[->, bend right=30] (chain2) edge (merge2);
      \draw[decorate, decoration={brace, mirror, amplitude=5pt, raise=5mm}]
          (x-conv2.south west) --
          node[label={[label distance=6mm] below:$1 \times 1$}] {}
          node[label={[label distance=9mm] below:kernel}] {}
          (x-conv2.south east);
      \draw[decorate, decoration={brace, mirror, amplitude=3pt, raise=3mm}]
          (chain1.north west) -- (chain1.south west);
      \draw[decorate, decoration={brace, mirror, amplitude=3pt, raise=3mm}]
          (chain2.north west) -- (chain2.south west);
    \end{tikzpicture}
    \caption{\textit{X-chain}, featuring cross-connections between the reference and
      difference image channels.
      Cross-connections and merging may occur before or after the BN and
      dropout layers.
    }
    \label{fig-x-chain}
  \end{center}
\end{figure}
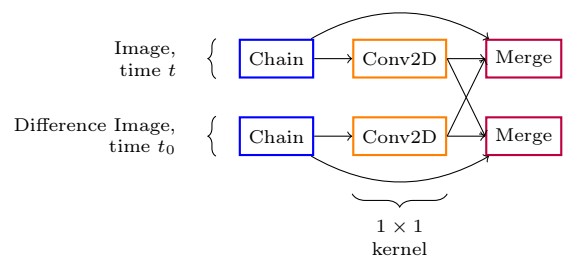

The basic building block of the convolutional part of each network is the \emph{chain},
shown for non-cross modal networks in Figure \ref{fig-chain}
and for cross-modal networks in Figure \ref{fig-x-chain}.
Chains consist of one or more convolutional layers, followed by pooling
and any optional regularisation layers - in this case, batch normalisation and dropout.
In the fully-connected layers, it was found that batch normalisation and dropout
conflicted, with batch normalisation on its own providing the best results.
Increasing the number of convolutional chains also proved to be unsuccessful,
with classification performance dropping as more chains were added.
This suggests the features of use to the network are subtle and that
further compression of them into more compact features by pooling loses information.
This supports the decision to not downsample the images during preprocessing.
Furthermore, it was found that increasing the number of neurons in the first
perceptron layer to 128 or 256 did not improve the models.
This was in fact advantageous, as the majority of trainable parameters in each model,
~$96\%$ for the differencing approaches,
came from the connections between the convolutional and perceptron layers.
Using fewer neurons in the first perceptron layer made the networks far smaller,
allowing faster training.

\newcommand{\chainhorizsep}{5mm}
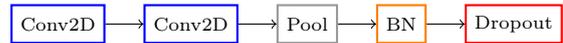
\begin{figure}[!t]
  \begin{center}
    \begin{tikzpicture}[
      every node/.style={font=\scriptsize},
      conv2d/.style={rectangle, draw=blue!100, fill=none, thick, minimum size=5mm,
      },
      pool/.style={rectangle, draw=black!40, fill=none, thick, minimum size=5mm,
      },
      bn/.style={rectangle, draw=orange, fill=none, thick, minimum size=5mm,
      },
      drop/.style={rectangle, draw=red, fill=none, thick, minimum size=5mm,
      },
      dense/.style={rectangle, draw=black, fill=none, thick, minimum size=5mm,
      },
    ]
      \node[conv2d] (conv1)                       {Conv2D};
      \node[conv2d] (conv2)   [right=\chainhorizsep of conv1]    {Conv2D};
      \node[pool]   (pool)    [right=\chainhorizsep of conv2]    {Pool};
      \node[bn]     (bn)      [right=\chainhorizsep of pool]     {BN};
      \node[drop]   (drop)    [right=\chainhorizsep of bn]       {Dropout};
      \path[->]     (conv1)   edge  (conv2);
      \path[->]     (conv2)   edge  (pool);
      \path[->]     (pool)    edge  (bn);
      \path[->]     (bn)   edge  (drop);
    \end{tikzpicture}
  \end{center}
  \caption{\emph{Chain} - a CNN building block:
    2 convolutional layers followed by max-pooling and
    regularised by batch normalisation and dropout.
  }
  \label{fig-chain}
\end{figure}
In the case of X-CNNs, cross-connections are appended to the chain structure
from Figure \ref{fig-chain}, creating what we refer to as \emph{X-chains},
which allow the exchange of feature maps between channels.
These cross-connections consist of $1 \times 1$ convolutions of each channel,
merged via tensor concatenation.
Regularisation layers such as batch normalisation may happen either before or
after the cross-connection process;
the impact of this is not yet fully understood.
%
%
%
%
\newcommand{\xcnnminwidth}{13mm}
\newcommand{\xcnntextminheight}{2em}
\newcommand{\xcnnhorizsep}{3mm}
\newcommand{\xcnndensevertsep}{2mm}
\newcommand{\xcnndensehorizsep}{7mm}
\begin{figure*}[!t]
  \begin{center}
    \begin{adjustbox}{width=\textwidth}
      \begin{tikzpicture}[
        every node/.style={
          font=\scriptsize,
          anchor=mid,
          text centered,
        },
        chain/.style={
          rectangle,
          draw=blue!100,
          fill=none,
          thick,
          text centered,
          minimum height=\xcnntextminheight,
          minimum width=\xcnnminwidth,
        },
        bn/.style={
          rectangle,
          draw=orange,
          fill=none,
          thick,
          minimum height=\xcnntextminheight,
          minimum width=\xcnnminwidth,
        },
        drop/.style={
          rectangle,
          draw=red,
          fill=none,
          thick,
          minimum height=\xcnntextminheight,
          minimum width=\xcnnminwidth,
        },
        input/.style={
          rectangle,
          draw=black!40,
          fill=none,
          thick,
          minimum height=\xcnntextminheight,
          minimum width=\xcnnminwidth,
        },
        output/.style={
          rectangle,
          draw=red,
          fill=none,
          thick,
          minimum height=\xcnntextminheight,
          minimum width=\xcnnminwidth,
        },
        dense/.style={
          rectangle,
          draw=black,
          fill=none,
          thick,
          minimum height=\xcnntextminheight,
          minimum width=\xcnnminwidth,
        },
        merge/.style={
          rectangle,
          draw=purple,
          fill=none,
          thick,
          minimum height=\xcnntextminheight,
          minimum width=\xcnnminwidth,
        },
        chainb/.style={rectangle, draw=blue, fill=blue,
          minimum height=10mm, minimum width=\xcnnminwidth,
        },
        bnb/.style={rectangle, draw=orange, fill=orange,
          minimum height=10mm, minimum width=\xcnnminwidth,
        },
        dropb/.style={rectangle, draw=red, fill=red,
          minimum height=10mm, minimum width=\xcnnminwidth,
        },
        mergeb/.style={rectangle, draw=purple, fill=purple,
          minimum height=25mm, minimum width=\xcnnminwidth,
        },
        inb/.style={rectangle, draw=black!40, fill=black!40,
          minimum height=10mm, minimum width=\xcnnminwidth,
        },
        outd/.style={circle, draw=red, fill=red,
          minimum size=1mm,
        },
        densed/.style={circle, draw=black, fill=black, minimum size=1mm,
        }
        ]
        \node[input]      (inp)                               {Input};
        \node[chain]      (chain)   [right=\xcnnhorizsep of inp]       {X-Chain};
        \node[merge]      (merge)   [right=\xcnnhorizsep of chain]     {Merge};
        \node[dense]      (dense1)  [right=\xcnnhorizsep of merge]     {Dense};
        \node[bn]         (bn1)     [right=\xcnnhorizsep of dense1]    {BN};
        \node[dense]      (dense2)  [right=\xcnnhorizsep of bn1]       {Dense};
        \node[bn]         (bn2)     [right=\xcnnhorizsep of dense2]    {BN};
        \node[output]     (out)     [right=\xcnnhorizsep of bn2]       {Output};
        \node[inb] (inbi) [below=3mm of inp] {};
        \node[inb] (inbd) [below=3mm of inbi] {};
        \node[inb] (inbt) [below=3mm of inbd] {};
        \node[chainb] (cbi) [right=\xcnnhorizsep of inbi] {};
        \node[chainb] (cbd) [right=\xcnnhorizsep of inbd] {};
        \node[mergeb] (mb1) [right=\xcnnhorizsep of cbd] {};
        \node[densed] (d13) [right=\xcnnhorizsep of mb1] {};
        \node[densed] (d12) [above=\xcnndensevertsep of d13] {};
        \node[densed] (d11) [above=\xcnndensevertsep of d12] {};
        \node[densed] (d14) [below=\xcnndensevertsep of d13] {};
        \node[densed] (d15) [below=\xcnndensevertsep of d14] {};
        \node[densed] (d23) [right=\xcnndensehorizsep of d13] {};
        \node[densed] (d22) [above=\xcnndensevertsep of d23] {};
        \node[densed] (d21) [above=\xcnndensevertsep of d22] {};
        \node[densed] (d24) [below=\xcnndensevertsep of d23] {};
        \node[densed] (d25) [below=\xcnndensevertsep of d24] {};
        \node[bnb] (bnb1) [right=\xcnnhorizsep of d23] {};
        \node[densed] (d33) [right=\xcnnhorizsep of bnb1] {};
        \node[densed] (d32) [above=\xcnndensevertsep of d33] {};
        \node[densed] (d31) [above=\xcnndensevertsep of d32] {};
        \node[densed] (d34) [below=\xcnndensevertsep of d33] {};
        \node[densed] (d35) [below=\xcnndensevertsep of d34] {};
        \node[densed] (d43) [right=\xcnndensehorizsep of d33] {};
        \node[densed] (d42) [above=\xcnndensevertsep of d43] {};
        \node[densed] (d41) [above=\xcnndensevertsep of d42] {};
        \node[densed] (d44) [below=\xcnndensevertsep of d43] {};
        \node[densed] (d45) [below=\xcnndensevertsep of d44] {};
        \node[bnb] (bnb2) [right=\xcnnhorizsep of d43] {};
        \node[densed] (d53) [right=\xcnnhorizsep of bnb2] {};
        \node[densed] (d52) [above=\xcnndensevertsep of d53] {};
        \node[densed] (d51) [above=\xcnndensevertsep of d52] {};
        \node[densed] (d54) [below=\xcnndensevertsep of d53] {};
        \node[densed] (d55) [below=\xcnndensevertsep of d54] {};
        \node[outd] (d62) [right=\xcnndensehorizsep of d53] {};
        \node[outd] (d61) [above=\xcnndensevertsep of d62] {};
        \node[outd] (d63) [below=\xcnndensevertsep of d62] {};
        \node[circle, minimum size=1mm, draw=white, fill=none]
                          (blank1) [right=\xcnnhorizsep of d61] {};
        \node             (blank2) [right=\xcnnhorizsep of d62] {};
        \node             (blank3) [right=\xcnnhorizsep of d63] {};
        \path[->]         (inbi)    edge    (cbi);
        \path[->]         (inbd)    edge    (cbd);
        \path[->]         (inbt)    edge    (mb1.south west);
        \path[dashed, -]  (cbi.north east)  edge    (mb1.north west);
        \path[dashed, -]  (cbi.south east)  edge    (mb1.west);
        \path[dashed, -]  (cbd.north east)  edge    (mb1.west);
        \path[dashed, -]  (cbd.south east)  edge    (mb1.south west);
        \path[dashed, -]  (mb1.north east)  edge    (d11.north);
        \path[dashed, -]  (mb1.south east)  edge    (d15.south);
        \path[->]         (d11) edge (d21)  (d12) edge  (d21) (d13) edge  (d21)
                          (d14) edge (d21)  (d15) edge  (d21);
        \path[->]         (d11) edge (d22)  (d12) edge  (d22) (d13) edge  (d22)
                          (d14) edge (d22)  (d15) edge  (d22);
        \path[->]         (d11) edge (d23)  (d12) edge  (d23) (d13) edge  (d23)
                          (d14) edge (d23)  (d15) edge  (d23);
        \path[->]         (d11) edge (d24)  (d12) edge  (d24) (d13) edge  (d24)
                          (d14) edge (d24)  (d15) edge  (d24);
        \path[->]         (d11) edge (d25)  (d12) edge  (d25) (d13) edge  (d25)
                          (d14) edge (d25)  (d15) edge  (d25);
        \path[dashed, -]  (d21.east)     edge    (bnb1.north west);
        \path[dashed, -]  (d25.east)     edge    (bnb1.south west);
        \path[dashed, -]  (bnb1.north east) edge    (d31.west);
        \path[dashed, -]  (bnb1.south east) edge    (d35.west);
        \path[->]         (d31) edge (d41) (d31) edge (d42) (d31) edge (d43)
                          (d31) edge (d44) (d31) edge (d45);
        \path[->]         (d32) edge (d41) (d32) edge (d42) (d32) edge (d43)
                          (d32) edge (d44) (d32) edge (d45);
        \path[->]         (d33) edge (d41) (d33) edge (d42) (d33) edge (d43)
                          (d33) edge (d44) (d33) edge (d45);
        \path[->]         (d34) edge (d41) (d34) edge (d42) (d34) edge (d43)
                          (d34) edge (d44) (d34) edge (d45);
        \path[->]         (d35) edge (d41) (d35) edge (d42) (d35) edge (d43)
                          (d35) edge (d44) (d35) edge (d45);
        \path[dashed, -]  (d41.east)     edge    (bnb2.north west);
        \path[dashed, -]  (d45.east)     edge    (bnb2.south west);
        \path[dashed, -]  (bnb2.north east) edge  (d51.west);
        \path[dashed, -]  (bnb2.south east) edge  (d55.west);
        \path[->]         (d51) edge (d61) (d51) edge (d62) (d51) edge (d63);
        \path[->]         (d52) edge (d61) (d52) edge (d62) (d52) edge (d63);
        \path[->]         (d53) edge (d61) (d53) edge (d62) (d53) edge (d63);
        \path[->]         (d54) edge (d61) (d54) edge (d62) (d54) edge (d63);
        \path[->]         (d55) edge (d61) (d55) edge (d62) (d55) edge (d63);
        \path[->]         (d61) edge (blank1);
        \path[->]         (d62) edge (blank2);
        \path[->]         (d63) edge (blank3);
        \draw[decorate, decoration={brace, mirror, amplitude=5pt, raise=42mm}]
            (chain.south west) --
            node[label={[label distance=42mm] below:$\times 5$}] {}
            (chain.south east);
        \draw[decorate, decoration={brace, mirror, amplitude=5pt, raise=42mm}]
            (dense1.south west) --
            node[align=center, label={[label distance=42mm] below:FC}] {}
            node[align=center, label={[label distance=45mm] below:64 neurons}] {}
            (dense1.south east);
        \draw[decorate, decoration={brace, mirror, amplitude=5pt, raise=42mm}]
            (dense2.south west) --
            node[label={[label distance=42mm] below:FC}] {}
            node[align=center, label={[label distance=45mm] below:32 neurons}] {}
            (dense2.south east);
        \draw[decorate, decoration={brace, mirror, amplitude=5pt, raise=42mm}]
            (out.south west) --
            node[label={[label distance=42mm] below:FC}] {}
            node[align=center, label={[label distance=45.6mm] below:$n$ neurons}] {}
            (out.south east);
        \draw[decorate, rotate=10, decoration={brace, amplitude=3pt, raise=2mm}]
            (d61.north) --
            node[label={[label distance=4mm] above:Softmax}] {}
            (blank1.north);
        \draw[decorate, decoration={brace, mirror, amplitude=5pt, raise=2mm}]
            (inbi.north west) --
            node[label={[label distance=4mm, rotate=0] left:Image, time $t$}] {}
            (inbi.south west);
        \draw[decorate, decoration={brace, mirror, amplitude=5pt, raise=2mm}]
            (inbd.north west) --
            node[label={[label distance=4mm, rotate=0] left:Difference Image, time $t_0$}] {}
            (inbd.south west);
        \draw[decorate, decoration={brace, mirror, amplitude=5pt, raise=2mm}]
            (inbt.north west) --
            node[label={[label distance=4mm, rotate=0] left:Timestamp $t$}] {}
            (inbt.south west);
      \end{tikzpicture}
    \end{adjustbox}
    \caption{X-CNN with channels for image data and its corresponding
      difference image data.
      Both channels are fed through five cross-connected chains, as in
      Figure \ref{fig-x-chain}.
      These and a timestamp channel are concatenated
      before being passed through fully-connected (FC) perceptron layers with batch normalisation.
      Softmax activations normalise the network output to the range $[0,1]$
      for each of the $n$ classes.
    }
    \label{fig-xcnn-t-diff}
  \end{center}
\end{figure*}
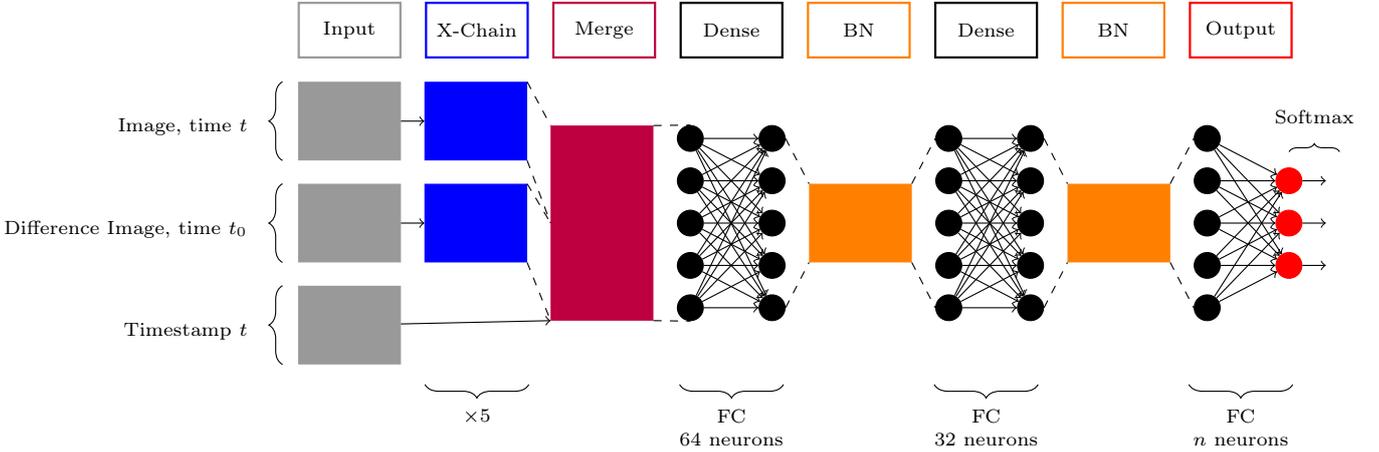
\vspace{-2em}
%
%
\subsection{Baseline CNN}
The baseline we used for comparison was a CNN, incorporating state-of-the-art
techniques.
There were five convolutional chains, the last of which did not use dropout
and fed to a series of fully-connected, ``dense'' layers,
comprising 64, 32, then $n$ neurons,
where $n$ was the number of treatment classes -- 2 in this case study.
All convolutions used $3 \times 3$ kernels -- 16 such kernels for the first chain
and 32 for each subsequent one.
The first two fully-connected layers were regularised by batch normalisation
on their outputs.
The max-pooling layers in the chains all used $2 \times 2$ non-overlapping blocks,
the dropout probability was $0.5$,
and $\gamma = 0.003$ for L2 regularisation in the fully-connected layers.
As mentioned, kernel initialisations for all but the output layer were
He Gaussians, accompanied by ReLU activations.
%
%
\subsection{Timestamps}
The most obvious and simple approach to incorporating time is the use of a
single number indicating the timestamp, which may be cardinal or ordinal.
As has been discussed, timestamps were not normalised.
The timestamp is provided via a unit-sized tensor, concatenated with the output
of the convolutional layers as input to the first fully-connected layer.
Otherwise, this model was identical to the baseline.
\subsection{Difference Images}
\label{sec-diff-images}
Difference images provide an implicit description of time by performing
a pixel-wise subtraction between an image at the time step, $t$, of interest ---
the \emph{comparison} image --- and an image at a preceding time step $t_0$ ---
the \emph{reference} image, should one exist:
\begin{align}
  D_{ij} = I^{(t)}_{ij} - I^{(t_0)}_{ij}
\end{align}
Two approaches to selecting the reference image are to use images from the
first recorded timestamp for each mouse or individual,
and to use images from the most recent timestamp available for each mouse
or individual.
These are referred to as \emph{absolute} and \emph{relative} differencing
respectively, as the timestamps selected are either absolute or are relative
to the current timestamp.
For either approach, the first timestamp's difference images will always be
of uniform intensity equal to a minimum value;
this represents the lack of preceding data.
That minimum value may be known in advance, e.g. $0$, or may be the minimum
value observed in each comparison image.
The former requires pre-calculation or determination of a minimum value,
which may not always be possible,
and may introduce values in the difference image much larger than would be
obtained in later timestamps' difference images,
placing undue emphasis on the first timestamp and the
least meaningful difference image.
Furthermore, it is inconsistent with the padding approach used in the image
expansion process in section \ref{sec-preprocessing}.
The latter approach, however, introduces global variation by allowing different
intitial-timestamp difference images to have different values.
We selected the latter approach as it is more suitable for large and real-time
data sets.
Differencing faced one further difficulty, in that for each mouse
there were a different number of images each week.
As such, a one-to-one matching between reference and comparison images was
not possible.
As each \textmu CT scan began at the proximal growth plate,
we paired the $k^{th}$ reference image with the $k^{th}$ comparison image
from the appropriate week,
stopping when there were no more images from either of the weeks.
This lead to $2.18\%$ of images being unused.

Having both original and difference images provides multiple processing channels
thereby enabling the use of cross-connections, making such models X-CNNs.
Processing was identical for absolute and relative differencing so that any
observed difference in performance could be attributed to the type of
differencing used, rather than potentially being the result of
different architectures or hyperparameters.
The processing in both convolutional channels used the same structure and hyperparameters,
namely five X-chains as in Figure \ref{fig-x-chain}
with dropout probability set to $0.25$,
He Gaussian kernel initialisations with ReLU activations,
and max-pooling over non-overlapping blocks with stride length $2$.
The number of convolution kernels used in each channel was eight for the first two chains,
16 for the subsequent two chains, and 32 for the final one.
The first two fully-connected perceptron layers contained 64 and 32 neurons respectively,
both using L2 regularisation with $\lambda = 0.0003$.
The final, output, perceptron layer was identical to that used in the other models.
%
%
%
%
\begin{table*}[!ht]
  \small
  \begin{center}
    \begin{tabular}{ | >{\itshape} c | *{5}{>{$} c <{$}|} }
      \hline
      \multirow{2}{*}{\normalfont \textbf{Method}} &
          \multicolumn{2}{ c | }{\bfseries Accuracy / \%} &
          \multicolumn{2}{ c | }{\bfseries Mean Time / s} &
          \multirow{2}{*}{\normalfont \textbf{Parameters}} \\
      \cline{2-5}
      & \mathrm{\textbf{5\ Epochs}} & \mathrm{\textbf{10\ Epochs}}
      & \mathrm{\textbf{5\ Epochs}} & \mathrm{\textbf{10\ Epochs}} & \\
      \hline
      CNN & 73.02 & 66.08 & 9677 & 9676 & 863883 \\ 
      \hline
      CNN, timestamps & 79.54 & 68.19 & 9631 & 9621 & 863947 \\ 
      \hline
      X-CNN, abs. diff. & 97.99 & 95.07 & 42033 & 44767 & 1643323 \\
      \hline
      X-CNN, rel. diff. & 84.28 & 85.36 & 42334 & 46816 & 1643323 \\
      \hline
      X-CNN, timestamps \& abs. diff. & 98.39 & 99.54 & 44196 & 43560 & 1643387 \\
      \hline
      X-CNN, timestamps \& rel. diff. & 83.43 & 75.15 & 43990 & 45356 & 1643387 \\
      \hline
    \end{tabular}
    \caption{Accuracy scores and per-epoch training times for
      the different neural network classifiers, after 5 epochs and 10 epochs of training,
      together with the number of parameters in each model.
    }
    \label{table-results}
  \end{center}
\end{table*}
\begin{table}[!hb]
  \small
  \begin{center}
    \begin{tabular}{| >{\bfseries} c | >{\bfseries} c | c | c | c | }
      \hline
      & & {\bfseries ``Wild'' / \%} & {\bfseries PTH / \%} & {\bfseries Total} \\
      \hline
      \multirow{3}{*}{No diff.}& Train. & 49.35 & 50.65 & 76663 \\
      \cline{2-5}
                                      & Val. & 49.70 & 50.30 & 8519 \\
      \cline{2-5}
                                      & Test & 50.24 & 49.76 & 21932 \\
      \hline
      \multirow{3}{*}{Diff.}   & Train. & 49.46 & 50.54 & 75004 \\
      \cline{2-5}
                                      & Val. & 49.34 & 50.66 & 8334 \\
      \cline{2-5}
                                      & Test & 50.04 & 49.96 & 21400 \\
      \hline
    \end{tabular}
    \caption{Percentages of observations in the ``wild type'' and PTH treatment classes
      and total number of observations, for the training, validation, and test sets,
      both when using image differencing and when using original images only.
    }
    \label{table-class-splits}
  \end{center}
\end{table}
%
%
%
%
\subsection{Combining Descriptions of Time}
It is, of course, possible to combine both explicit descriptions of time,
like timestamps, with implicit ones, like differencing.
Providing a network with both representations allows it to determine the
relative importance of each during training,
thereby adapting better to the particular domain and problem.
There may be situations where one temporal descriptor is more useful than another
or when one can be a deciding factor.

For combined temporal descriptions, we have three separate input channels:
the original image, the difference image, and the timestamp.
The first two undergo the same processing as in section \ref{sec-diff-images},
then the tensors from all three channels are concatenated before input to the
fully-connected perceptron layers.
This is shown in Figure \ref{fig-xcnn-t-diff}.
All hyperparameter choices were as in section \ref{sec-diff-images}.
%
%
%
%
\section{Results}
In Table \ref{table-results}, we present the test-set accuracies and
mean training times over five and 10 epochs,
together with the number of parameters (both trainable and non-trainable),
for each of the six models considered.
We use accuracy as an evaluation metric
as the training, validation, and test sets were all almost perfectly
balanced in terms of the number of images available for each treatment class.
The exact splits are shown in Table \ref{table-class-splits}.
We provide the average training time per epoch, in addition to the
number of epochs of training, as real-world training times for new models
may impact decisions on which models to use in a medical setting
based on available resources and how frequently models need to be re-trained.

The best performing models were those using absolute differencing,
achieving over $95\%$ accuracy after both 5 and 10 epochs of training,
both with and without timestamps.
The inclusion of timestamps with absolute differencing yielded the best
results overall - $99.54\%$ accuracy after 10 epochs.
In contrast, the CNN baseline managed only $73.02\%$ accuracy,
achieved with 5 epochs of training.
Whilst not out of line with the performance of many
classification and prediction tasks on medical datasets
\cite{ashinsky17, guler05, rez16, seera17},
the relatively poor performance of the baseline when compared to commonplace results
on datasets such as MNIST and CIFAR-10 highlights the inherent difficulty of medical tasks
and the necessity of finding novel, informative features.
We see that, for the same number of training epochs,
spatio-temporal models always outperformed the purely spatial baseline
and that only in one of the 10 results for the temporal models
did the accuracy fall below a baseline result:
the CNN with timestamps, trained for 10 epochs,
performed worse than the baseline CNN trained for five epochs.
In general, it was the case that training for longer, 10 rather than five epochs,
offered little to no benefit ---
for only two of the six models was there a performance increase, and then only of circa $1\%$,
whereas the other models saw drops of around $3-11\%$, averaging a $-7.37\%$ decrease.
This suggests the models may be susceptible to overfitting,
at least on a deep rather than wide dataset as in the case study,
despite extensive use of regularisation techniques.
Given that performance after only five epochs could reach around $98\%$,
this is in fact positive -- powerful, useful models can be trained more quickly.
Note that if difference images provided no information beyond
that contained in the original images,
we would expect to find that training the non-differencing models for twice as many epochs
would allow those models to reach the same accuracies as with differencing,
as they would have processed the same total quantity of information.
As this was not the case, we must conclude that difference images contain
valuable additional information.

Whilst relative differencing exceeded the baseline and timestamp accuracies,
it performed noticeably worse than absolute differencing ---
between approximately $10-25\%$ worse, depending on the number of training epochs
and the inclusion or exclusion of timestamps.
This is likely a consequence of the dependency range the model can represent,
with relative differencing only capturing short-range dependencies in a limited temporal region,
wheras absolute differencing is capable of modelling arbitrarily long-range dependencies
as RNNs can, although more efficiently as RNNs build long-range dependencies from
iterating over short-range ones.
For conditions like abnormal bone remodelling which,
when averaged over cycles of periodic behaviour,
exhibit monotonic trends in a metric of interest,
such as bone density or volume,
long-range dependencies may best capture these trends.
Relative differencing may be more suitable for conditions characterised by
fluctuations around a central (range of) value(s),
which may average out over longer time spans.
The efficacy of timestamps may be related to this,
in that for absolute differencing, timestamps reinforce the information provided
by the difference images - larger pixel values, indicating larger differences in
bone density, tend to correlate with larger timestamps.
On the other hand, relative differencing may produce similar difference images
for each week, thereby making it difficult to distinguish which week a difference
image was from; timestamps would not correlate with the pixel intensities in the
difference images, thus attempting to coordinate the information from these two
temporal descriptions may produce erroneous associations.
This would explain the impaired performance when using relative differencing
with timestamps.

From Table \ref{table-results}, we see that the non-cross modal models required
fewer than 900K parameters each, whilst the cross-modal ones required a little over
1.5 million each.
In comparison with recent high-performance image classification networks ---
5 million parameters in GoogLeNet, 23 million in Inception v3, 60 million in AlexNet,
and 180 million in VGGNet \cite{szegedy15} ---
a 1--1.5 million parameter network is computationally inexpensive.
Based on the training times in Table \ref{table-results}, we see two main points.
Firstly, that the difference image-based approaches take around four times as long as
the non-differencing ones, likely due to a combination of disk-load speeds from
a larger-than-memory dataset and the cost of training a more complex network with
around twice as many parameters.
Secondly, training these networks necessitates the use of GPUs or highly parallel
multi-core processing systems --- whilst smaller than many high-performance models,
they are not sufficiently computationally inexpensive on train on standard desktop systems.
%
%
\section{Conclusions}

We have shown several approaches to the incorporation of temporal information
for image-processing tasks and compared their performance on a balanced dataset
comprising over 100 000 data points across two treatment groups.
For the same number of epochs of training, the models with temporal descriptions
always outperformed an atemporal state-of-the-art CNN baseline,
in some cases by in excess of $25\%$,
highlighting the distinct potential to improve atemporal models of
temporally-varying processes and systems.
Whilst even simple mechanisms for providing temporal information can yield benefits,
such as integer timestamps,
the best performance is achieved by using a temporal description suitable for the
domain and problem.
For our case study, wherein the treatments produced monotonic effects
over extended periods of time, the ability of absolute differencing to
capture long-range dependencies produced the best results.
For certain medical applications this is highly advantageous ---
accurately determining the disease or treatment status of a patient
may require as little as two scans some number of weeks apart,
without the need for invasive procedures or regular appointments.
Furthermore, the temporal models we present permit sparse, non-sequential data
sampled at irregular intervals, making them far more flexible than RNNs and 3D CNNs
and without the need for interpolation or other approaches to reconstruct
sequential data, which ultimately add a preprocessing cost for those types of networks.

Whilst we have shown the effectiveness of cross-modal neural networks for
a particular temporally-varying medical dataset,
our approach was kept general throughout and is applicable to a wider medical context.
The models we developed could easily be extended or modified to particular domains
or for further research into alternative temporal descriptions.
Potential extensions include the use of both absolute and relative differencing,
to capture both long- and short-range dependencies,
the use of SVMs or RNNs either in place of or subsequent to the fully-connected layers,
and the application of our models to colour images.
The latter would allow investigation of whether differencing extracts additional
information from colour channels or only from the original image or luminance channel.
In order to create less computationally-expensive and resource-intensive networks,
the use of only difference images without their corresponding reference images should
be explored, as they retain some amount of spatial information whilst representing
temporal variations.

The proposed approach has the potential for improving the current assessment of
the effects of treatments for musculoskeletal pathologies preclinically.
The classification of the significant effects of a treatment can be used to
optimise the dosage, time of treatment, or even combined interventions,
for example alternate or overlapped treatment with anabolic and anti-resorptive drugs,
or usage in combination with mechanical stimulation.
Longitudinal experiments in mice are expensive and have ethical issues
related to the usage of animals in research.
Therefore, improving the classification of the effects of interventions
with an automated, operator-insensitive tool would lead to a dramatic reduction in
cost and the number of animals in research, in line with the 3 Rs:
reduction, refinement, and replacement.

From the perspective of the longitudinal characterisation of phenotypes,
we believe that there are two main challenges for the future.
Firstly, if researchers wish to have broad oversight on the effects of
ageing, co-mordibities, and their related interventions
rather than focusing on specific tissues, organs, or systems,
then it will be vital to combine data obtained from different modalities.
As the data collected from different modalities is unlikely to be independent,
uncovering powerful, effective ways to combine this information should result
in models which are greater than the sum of their parts.
Differencing goes some way towards this by using a spatial format --- an image ---
to represent a change over time,
and cross-modal networks support feature-sharing between selected modalities,
allowing researchers to encode domain-specific knowledge via carefully chosen
cross-connections.
The second challenge we perceive is the need for increased image resolution
to capture features which are not presently available,
such as osteocyte lacunae within the extracellular matrix.
The availability of such high-resolution images and the features they would unlock
would require research on how to create machine learning approaches capable of
accommodating the width, variety, and detail of such features in an efficient,
flexible manner, with consideration for the likelihood of data points being
sampled irregularly, at different rates, and with missing values.

%
%
\bibliographystyle{bib_style}
{\small \bibliography{citations}}
%
\end{document}